\documentclass{ws-ijgmmp}

\begin{document}

\markboth{D. Momeni et al.}
{Cosmological viable  Mimetic $f(R)$ and $f(R,T)$ theories via Noether symmetry}

%
\catchline{}{}{}{}{}
%

\title{Cosmological viable  Mimetic $f(R)$ and $f(R,T)$ theories via Noether symmetry}

\author{D. Momeni}

\address{Eurasian International Center for Theoretical Physics and Department of
General \& Theoretical Physics, Eurasian National University, \\
Astana 010008, Kazakhstan\\
\email{d.momeni@yahoo.com} }

\author{E. G\"{u}dekli}

\address{Department of Physics, Istanbul University, Istanbul,
Turkey }

\author{R. Myrzakulov}

\address{Eurasian International Center for Theoretical Physics and Department of
General \& Theoretical Physics, Eurasian National University, \\
Astana 010008, Kazakhstan\\
\email{rmyrzakulov@gmail.com} }

\maketitle

\begin{history}
\received{(Day Month Year)}
\revised{(Day Month Year)}
\end{history}

\begin{abstract}
Extended $f(R)$ theories of gravity have been investigated from the symmetry point of view. We briefly has been investigated Noether symmetry of two types of extended $f(R)$ theories: $f(R,T)$ theory, in which curvature is coupled non minimally to the trace of energy momentum tensor $T_{\mu\nu}$ and mimetic $f (R) $ gravity, a theory with a scalar field degree of freedom, but ghost-free and with internal conformal symmetry. In both cases we write point -like Lagrangian for  flat  Friedmann-Lemaitre-Robertson-Walker  (FLRW)  cosmological background in the presence of ordinary matter. We have been shown that some classes of models existed with Noether symmetry in these viable extensions of $f(R)$ gravity. As a motivated idea, we have been investigating the stability of the solutions and the  bouncing and $\Lambda$CDM models using  the  Noether symmetries. We have been shown that  in mimetic $f(R)$ gravity bouncing and $\Lambda$CDM solutions are possible. Also a class of solutions with future singularities has been investigated.
\end{abstract}


\section{Introduction}

A challenge to the contemporary relativistic cosmology is provided by a set of observational data, indicated on late time acceleration expansion of the whole Universe as well as the initial era, the inflationary epoch \cite{Ri98-1}-\cite{Ri98-4}. In the framework of the general relativity (GR), inflationary model, quasi stable de Sitter cannot be realized as a possible physically acceptable model without any extra scalar matter field. It is needed to include some  matter fields, like scalar field(s) to resolve it. Another approach which is totally revolutionary is to replace Einstein gravity by some extended forms of classical geometrical objects.  A reasonable candidate to explain this situation is modified gravity, in which we replace Einstein-Hilbert action, given by 
\begin{eqnarray}
S_{EH}=\int_{\mathcal{M}}{\frac{R}{2\kappa^2}\sqrt{-g}d^4x}+ S_{EH}|_{\partial{\mathcal{M}}}
\end{eqnarray}
 (Here $S_{EH}|_{\partial{\mathcal{M}}}$ is the boundary term) by another set of geometrical objects like second order invariants $R^{\mu\nu}R_{\mu\nu}, R^{\mu\nu\alpha\beta}R_{\mu\nu\alpha\beta},..$ or 
functions of Ricci scalar $R$, Gauss-Bonnet topological invariant $G=R^{\mu\nu\alpha\beta}R_{\mu\nu\alpha\beta}
-4R^{\mu\nu}R_{\mu\nu}+R^2$ and so on (see \cite{Nojiri:2010wj}-
\cite{ijgmmp} for reviews). Historically, to be more precise, the simplest potentially reasonable candidate was $f (R) $ gravity, a theory which it was proposed originally before recent activities \cite{Buchdahl}) and later motivated in light of the recent observational data, as a valid, physically reasonable, ghost-free and stable alternative theory instead of GR \cite{Nojiri:2010wj}-
\cite{ijgmmp}.  \par
Another type of modified gravity is the one in which geometry has been coupled to the matter fields non- minimally (see \cite{Lobo:2014ara} for
 An updated review of such models). Different types of non-minimally coupled models have been proposed like $f(R,\mathcal{L}_m)$ where $\mathcal{L}_m$ stands for matter Lagrangian and $f(R,T)$ \cite{Harko:2011kv},where $T$ is the  trace of the energy momentum tensor of matter fields,which is  defined by 
\begin{equation}\label{enmom}
T_{\mu \nu }=-\frac{2}{\sqrt{-g}}
\frac{\delta \left( \sqrt{-g}L_\mathrm{m}\right) }{\delta g^{\mu \nu }}\,,
\end{equation}
Or after a simple checking, it can be rewritten as the following\footnote{This common definition of energy-momentum tensor can not be used to fix the form of the Lagrangian. For example, of perfect fluid there is no way to read Lagrangian from the form of $T_{\mu\nu}$}:
\begin{equation}\label{en1}
T_{\mu \nu }=g_{\mu \nu }L_\mathrm{m}-2
\frac{\partial L_\mathrm{m}}{\partial g^{\mu\nu}}\,.
\end{equation}
In this theory, $T=T_{\mu}^{\mu}$. Because of simplicity and beauty form, this theory attracted several activities in literature  \cite{Momeni:2011am}-\cite{Singh:2014bha}. 
Symmetry is an important issue to be addressed in any physical system under study. There are two classes of symmetries: global symmetries , in which the physical system( dynamical system) respects some types of transformations, which are defined by functions of coordinates. Another is local, in which the conservation law gives us a "local" conserved quantity like charge.

As we mentioned it before, $f (R) $ gravity is a ghost free, and conformally equivalent to the scalar field theory in Einstein frame. There are several interesting features in this theory to be useful for late time acceleration, dark energy and dark matter halo problems. Also, inflation can be realized successfully using this simple and effective modified theory of gravity. Recently to resolve dark matter problem, a new type of modified gravities has been proposed as titled mimetic model and modified versions of it
\cite{Chamseddine:2013kea}-\cite{Momeni:2014qta}. Basic hidden idea behind this new modification of gravity is to propose a conformally invariant, scalar theory of gravity in which scalar degree of freedom does not cause any problem with ghosts. The way is to reparametrize metric as a conformal transformation of an auxiliary metric ,"unphysical" metric. The point is, the scalar field appeared as conformal function and plays the role of an internal degree of freedom. This scalar field is not ghost and it is constructed to have unit norm $\partial_{\mu}\phi\partial^{\mu}\phi=-1$. This norm is defined on physical metric. It is remembering for us the role of velocity of a test particle with unit norm in the comoving frame of particle coordinates. An interesting feature is if we write FLRW cosmological equations, 
 an extra term proportional 
to $a^ {-3} $, appeared. It mimics dark matter. So, it was adequate to name it as  ``mimetic dark matter'' or briefly mimetic gravity. To unify $f(R)$ gravity with this very interesting mimetic theory, it has been proposed mimetic $f(R)$ gravity 
\cite{Nojiri:2014zqa} as 
a new class of modified gravities with the same inspiration as mimetic theory. Because of its complexity and more physical solutions, this new mimetic $f(R)$   deserves further physical investigations. Very recently the dynamical behavior of mimetic f (R) has been investigated 
 \cite{Leon:2014yua}. Our aim here is to address Noether symmetry issue of such mimetic models. 
In literature Noether symmetry has been investigated for different types of modified gravity like $f(R)$, $f(T)$, Galileons and so on \cite{Capozziello:2012hm}-\cite{Jamil:2011pv}. Our aim in this paper is to explore Noether symmetry for the mentioned above to modify gravities: mimetic $f (R)$ and $f (R, T)$.
\par
The present paper is organized as follows: formal framework for $f(R,T)$ theory of gravity  is presented in Section~\ref{f(R,T)}. The formalism of f (R) mimetic theory is motivated in
Section~\ref{f(R)mimetic}. The Noether symmetry approach is well understood briefly in
Section~\ref{NS}. Noether symmetry is applied to $f(R,T)$ in Section~\ref{NS1} and for $f(R)$ mimetic model
in Section~\ref{NS2}. We summarize in the last section.

 We adopted a "God" given system of units $G=c=1$, where the  gravitational
coupling constant is given by $\kappa^2 =8\pi$.

\section{Formal framework of  $f\left(R,T\right)$
gravity}\label{f(R,T)}
Let us to start by the following simple extension of $f(R)$ gravity in four dimensional spacetime:
\begin{equation}
S=\frac{1}{16\pi}\int
f\left(R,T\right)\sqrt{-g}\;d^{4}x+\int{L_\mathrm{m}\sqrt{-g}\;d^{4}x}\, ,
\end{equation}
In the action, $R$ stands for the Ricci scalar of the Riemannian space-time, $T$ is trace of the energy, momentum tensor of matter Lagrangian $L_\mathrm{m}$, is defined by the simple expression $T=g^{\mu\nu}T_{\mu\nu}$. Trace is an important quantity in quantum theory of inflation and quantum gravity. The dynamical quantity is just metric $g_{\mu\nu}$. So, there is only a single field equation for it. If we calculate the variation of action we obtain:
\begin{eqnarray}
\delta S&=&\frac{1}{16\pi }\int \left[ f_{R}\left( R,T\right)
\delta R+f_{T}\left(R,T\right) \frac{\delta T}{\delta g^{\mu \nu }}
\delta g^{\mu \nu }-\frac{1}{2}g_{\mu \nu }f\left( R,T\right) \delta g^{\mu
\nu }
+16\pi \frac{1}{\sqrt{-g}}
\frac{\delta \left( \sqrt{-g}L_\mathrm{m}\right) }{\delta g^{\mu \nu
}}\right]
\sqrt{-g}d^{4}x\, ,
\end{eqnarray}
In this variation, we define :  $f_{R}\left( R,T\right) =\partial
f\left(R,T\right) /\partial R$ and $f_T\left( R,T\right) =\partial
f\left(R,T\right) /\partial T$, respectively. 
Since $R$ is the only geometrical quantity in our theory, we recall standard variational expressions for it as the following:
\begin{equation}
\delta R=\delta \left( g^{\mu \nu }R_{\mu \nu }\right) =R_{\mu \nu }\delta
g^{\mu \nu }+g^{\mu \nu }\left( \nabla _{\lambda }
\delta \Gamma _{\mu \nu}^{\lambda }-\nabla _{\nu }\delta \Gamma _{\mu
\lambda }^{\lambda }\right)\, ,
\end{equation}
A simple check point is to evaluate it in a local flat coordinate frame, when $\Gamma_{\alpha\beta}^{\gamma}\equiv0$.  A variation of the connection termination, $\Gamma$ gives us:
\begin{equation}
\delta \Gamma _{\mu \nu }^{\lambda }=\frac{1}{2}g^{\lambda \alpha }\left(
\nabla _{\mu }\delta g_{\nu \alpha }+\nabla _{\nu }
\delta g_{\alpha \mu}-\nabla _{\alpha }\delta g_{\mu \nu }\right)\, ,
\end{equation}
Consequently, we can write the final simplified form of variation of $R$ as the following:
\begin{equation}
\delta R=R_{\mu \nu }\delta g^{\mu \nu }+g_{\mu \nu }\square
\delta g^{\mu\nu }-\nabla _{\mu }\nabla _{\nu }\delta g^{\mu \nu }\, .
\end{equation}
By plugging these expressions in the total variation of action, $\delta S$, we find:
\begin{eqnarray}
\delta S&=&\frac{1}{16\pi }\int \Big[ f_{R}\left( R,T\right) R_{\mu \nu }
\delta g^{\mu\nu }+f_{R}\left( R,T\right) g_{\mu \nu }\square
\delta g^{\mu \nu}-f_{R}\left( R,T\right) \nabla _{\mu }\nabla _{\nu }\delta
g^{\mu \nu}
\nonumber  \\
&& +f_{T}\left( R,T\right) \frac{\delta \left(g^{\alpha \beta }T_{\alpha
\beta }\right)}{\delta
g^{\mu \nu }} \delta g^{\mu \nu }-\frac{1}{2}g_{\mu \nu }f\left( R,T\right)
\delta
g^{\mu \nu }+16\pi \frac{1}{\sqrt{-g}}
\frac{\delta \left( \sqrt{-g}L_\mathrm{m}\right) }{\delta g^{\mu \nu
}}\Bigg]
\sqrt{-g}d^{4}x\, .  \label{var1}
\end{eqnarray}
To simplify more these functional , we need to redefine an auxiliary tensor field $\Theta _{\mu \nu}$ as a part of the variation of trace $\delta T$ as the following:
\begin{equation}
\frac{\delta \left(g^{\alpha \beta }T_{\alpha \beta }\right)}{\delta g^{\mu
\nu}}
=T_{\mu\nu}+\Theta _{\mu \nu}\, ,
\end{equation}
And in a similar form as we define $T$, we are able to define trace of $\Theta _{\mu \nu}$ as the following:
\begin{equation}
\Theta_{\mu \nu}\equiv g^{\alpha \beta }\frac{\delta T_{\alpha \beta
}}{\delta g^{\mu \nu}}\, .
\end{equation}
Using these simplifications, we write the following form of the equation of motion of $f(R,T)$:
\begin{eqnarray}\label{field}
f_{R}\left( R,T\right) R_{\mu \nu } - \frac{1}{2}
f\left( R,T\right)  g_{\mu \nu }
+\left( g_{\mu \nu }\square -\nabla_{\mu }\nabla _{\nu }\right)
f_{R}\left( R,T\right) =8\pi T_{\mu \nu}-f_{T}\left( R,T\right)
T_{\mu \nu }-f_T\left( R,T\right)\Theta _{\mu \nu}\, .
\end{eqnarray}
It is remarkable that, when geometry is decoupled from the matter part, in the limit  $f(R,T)\equiv f(R)$, we recover equation of motion of $f(R)$ gravity from (\ref{field}). Trace of (\ref{field}) gives us the following equation:

\begin{eqnarray}
&&f_{R}\left( R,T\right) R+3\square f_{R}\left( R,T\right) -2
f\left( R,T\right) =8\pi T-f_{T}\left(R,T\right)
T-f_{T}\left(R,T\right)\Theta\, ,
\label{contr}
\end{eqnarray}
An alternative form of (\ref{field}) is obtained as the following:
\begin{eqnarray}
&&f_{R}\left( R,T\right) \left( R_{\mu \nu }-\frac{1}{3}Rg_{\mu \nu
}\right) +\frac{1}{6} f\left( R,T\right)  g_{\mu\nu }
=8\pi \left(T_{\mu \nu}-\frac{1}{3}T
g_{\mu \nu}\right)-f_{T}\left( R,T\right)
\left( T_{\mu \nu }-\frac{1}{3}T g_{\mu \nu }\right)
\nonumber  \\
&&-f_T\left(R,T\right)\left(\Theta_{\mu \nu}-\frac{1}{3}
\Theta g_{\mu\nu}\right)
+ \nabla _{\mu }\nabla _{\nu }f_{R}\left(R,T\right)\, .
\end{eqnarray}
Conservation of the energy, momentum tensor in this type of modified gravity is checked by computing 
 the covariant divergence of Eq.~(\ref{field}). Using the following geometrical identity we obtain:
\begin{eqnarray}
\nabla ^{\mu }\left[ f_R\left(R,T\right)
R_{\mu\nu}-\frac{1}{2}f\left(R,T\right)g_{\mu\nu}
+\left(g_{\mu \nu }\square -\nabla_{\mu }\nabla_{\nu}\right)
f_R\left(R,T\right)\right] \equiv 0\, ,
\end{eqnarray}
So, conservation of energy, momentum tensor, gives us the following vector current term:
\begin{equation}\label{noncons}
\nabla ^{\mu }T_{\mu \nu }
\equiv J_{\nu}=\frac{f_{T}\left( R,T\right) }{8\pi -f_{T}\left(R,T\right) }
\left[ \left( T_{\mu \nu }+\Theta _{\mu \nu }\right) \nabla^{\mu }
\ln f_{T}\left( R,T\right) +\nabla ^{\mu }\Theta _{\mu \nu }\right]\, .
\end{equation}
For a simplicity, we consider  the flat   FLRW
metric in the following form:
\begin{equation}
ds^2= -dt^2+a^2(t)\,\delta_{ij} dx^i dx^j,\label{metric}
\end{equation}
Here  $a (t) $ stands for  the scale factor. If we write down Lagrangian of $f(R,T)$ for this metric and if we assumed that the Universe is filled with matter fields with effective pressure $p$ and energy density $\rho$, we obtain $T=3p-\rho$. Since this equation is a constraint, and because of $R=6(\frac{\ddot{a}}{a}+(\frac{\dot{a}}{a})^2)$, we introduce a pair of Lagrange multipliers as the following $\{\lambda,\mu\}$. The point Like Lagrangian after an integration part by part is written as the following (we set $2\kappa^2=1$):
\begin{eqnarray}
&&\mathcal{L}(a,R,T,\dot{a},\dot{R},\dot{T})=a^3\Big( f(R,T)-R f_{R}(R,T)-Tf_{T}(R,T)\Big)\\&&\nonumber-6\Big(a\dot{a}^2f_{R}(R,T)+a^2\dot{a}\dot{R}f_{RR}(R,T)+a^2\dot{a}\dot{T}f_{RT}(R,T)\Big)-p(a)a^3+a^3f_{T}\Big(3p(a)-\rho(a)\Big)\label{L1}
\end{eqnarray} 
Where we suppose that $\mathcal{L}_m=-p(a)$. If $f(R,T)=f(R)$, the Lagrangian density is written:
\begin{eqnarray}
\mathcal{L}(a,R,\dot{a},\dot{R})=a^3\Big( f(R)-R f_{R}(R)\Big)-6\Big(a\dot{a}^2f_{R}(R)+a^2\dot{a}\dot{R}f_{RR}(R)\Big)-p(a)a^3.
\end{eqnarray} 
This is the form of Lagrangian in $f(R)$ theory.
We suppose that (\ref{L1}) is an acceptable assumption for many cosmological applications, like matter dominant era or radiation. The associated equations of motion are given by a set of second order ordinary differential equations, Euler-Lagrange (EL) equations, are given by the following:
 \begin{eqnarray}
\frac{d}{dt}(\frac{\partial\mathcal{L}}{\partial \dot{q}})-\frac{\partial\mathcal{L}}{\partial q}=0
\end{eqnarray}
 for $q\equiv\{a,R,T\}$ they are obtained as the foloowing:
\begin{eqnarray}
&&6\ddot{R}f_{RR}=-3f  +3Rf _R +3Tf _T -6\,{\frac 
{{\dot{a}}^{2} }{{a_{{}}}^{2}}}f_{R}
-12\frac{\dot{a}\dot{R}}{a}f_{RR}-12\frac{\dot{a}\dot{T}}{a}f_{RT}+a_{{}}
p'(a)+3\,p \left( a_{{}}
 \right) -9f_{T} p \left( a_{{}} \right) \\&&\nonumber+3f_{T}
\rho \left( a_{{}} \right) -3\,a_{{}} f_{T} p'(a) +a_{{}} f_{T} \rho'(a) \left( a_{{}} \right) -6\dot{R}^{2}f_{RRR} -12\dot{R}
\dot{T}f_{RRT} -6\dot{T}^{2}f _{RTT} -
6f_{RT}\ddot{T}-12\frac{\ddot{a}}{a}f_{R}
\\&&
6f_{RR}\frac{\ddot{a}}{a}=f_{RR}\Big(R-6(\frac{\dot{a}}{a})^2\Big)+f_{RT}\Big(T-(3p(a)-\rho(a))\Big)
\\&& (R-6\frac{\dot{a}^2}{a^2}-6\frac{\ddot{a}}{a})f_{RT}+f_{TT}(T-3p(a)+\rho(a))=0.
\end{eqnarray}
To pass the case of $f(R,T)=f(R)$, we obtain:
\begin{eqnarray}
&&6\ddot{R}f_{RR}=-3f  +3Rf _R -6\,{\frac 
{{\dot{a}}^{2} }{{a_{{}}}^{2}}}f_{R}
-12\frac{\dot{a}\dot{R}}{a}f_{RR}+a_{{}}p'(a) +3\,p \left( a_{{}}
 \right) \\&&\nonumber -6\dot{R}^{2}f_{RRR}-12\frac{\ddot{a}}{a}f_{R}
\\&&
R=6\Big((\frac{\dot{a}}{a})^2+\frac{\ddot{a}}{a}\Big).
\end{eqnarray}

Our aim in this paper is to investigate Noether symmetry issue of (\ref{L1}). Briefly, we are interested to know how Noether symmetry is able to "fix" mathematical forms of $\{f(R,T),p(a)\}$.


\section{formalism of mimetic $f(R)$ gravity}\label{f(R)mimetic}
Although $f(R,T)$ provides a reasonable and good extension of $f(R)$ theory, it does not respect conformal symmetry. Also, extra degrees of freedom are possible. So, instabilities due to ghosts probably are happening. To resolve conformal symmetry and to be ghost-free, a model recently proposed as titled 
 Mimetic $F (R) $ gravity 
\cite{Nojiri:2014zqa}. It is inspired from the mimetic theory \cite{Chamseddine:2013kea,Golovnev:2013jxa,Chamseddine:2014vna,Chaichian:2014qba,
Malaeb:2014vua,
Deruelle:2014zza,Momeni:2014qta}, a model in which dark matter problem is resolved as an integration constant. Also, it is self consistent with conformal symmetry. The basis of any type of mimetic theory is to 
 parameterize of the Riemannian metric tensor  
 $g_{\mu\nu}$  as the following conformally transformed formula \cite{Chamseddine:2013kea}
\begin{equation}
\label{metricreparametrization}
g_{\mu\nu}= - \hat {g}^{\rho\sigma} \partial_\rho \phi \partial_\sigma \phi 
\hat {g}_{\mu\nu} \,,
\end{equation}
Here we introduced a pair of auxiliary objects: the first is an auxiliary metric (unphysical and without dynamics)
  $\hat {g}_{\mu\nu}$ and the second is a scalar field degree of freedom
$\phi$ which has generally ghost, freedom. It is well known that using an orthogonality of metric, this scalar field satisfies the following constraint equation of motion:
\begin{equation}
\label{phimetricrelation}
g\left({\hat g}_{\mu\nu}, \phi \right)^{\mu\nu} \partial_\mu \phi 
\partial_\nu \phi = - 1\, ,
\end{equation}
If we know the background metric $g\left({\hat g}_{\mu\nu}, \phi \right)^{\mu\nu}$, this equation fixes the form of the scalar field. If we interpret $\partial_\mu \phi $ as the components of a four velocity $u_{\mu}$, then normalized $u_{\mu}u^{\mu}=-1$ implies a possible normalization of the $\phi$. This normalization can be understood as the first integral of the equation of motion for 
 $\phi$ . We emphasize here that the auxiliary metric $\hat {g}_{\mu\nu}$ is an internal object of the space-time manifold 
Following  \cite{Nojiri:2014zqa}, we write the following action for mimetic $f(R)$ gravity in metric formalism :
\begin{equation}
\label{FRstandact}
S=\int d^4 x \sqrt{-g} \left[\frac{ f(R)}{2\kappa^2} 
+ \mathcal{L}_m\right]\, ,
\end{equation}
Here $\kappa^2=8\pi$, $R$ is the Ricci scalar which is computed by the 
physical metric $g_{\mu\nu}$, and we also include the matter 
Lagrangian by $\mathcal{L}_m$ . What we need is to 
parametrize the physical metric according to (\ref{metricreparametrization}) . We rewrite the action of theory in the following equivalent form:

\begin{equation}
\label{newFRaction}
S=\int d^4 x \sqrt{-g\left(\hat {g}_{\mu\nu}, \phi \right)}
\left[ \frac{f\left(R\left({\hat g}_{\mu\nu}, \phi \right)\right)}{2\kappa^2}
+ \mathcal{L}_m\right]\, .
\end{equation}
Thus, we perform variation with respect to auxiliary metric $\hat {g}_{\mu\nu}$ , we obtain:
\begin{eqnarray}
\label{Eq1}
 &&\frac{1}{2} g_{\mu\nu} f\left(R\left({\hat g}_{\mu\nu}, \phi \right)\right)
 - R\left({\hat g}_{\mu\nu}, \phi \right)_{\mu\nu}
f_R\left(R\left({\hat g}_{\mu\nu}, \phi \right)\right)
\nonumber\\
&&
+ \nabla\left(g\left({\hat g}_{\mu\nu}, \phi \right)_{\mu\nu}\right)_\mu
\nabla\left(g\left({\hat g}_{\mu\nu}, \phi \right)_{\mu\nu} \right)_\nu
f_R\left(R\left({\hat g}_{\mu\nu}, \phi \right)\right) 
\nonumber\\
&&
 - g\left({\hat g}_{\mu\nu}, \phi \right)_{\mu\nu}
\Box \left({\hat g}_{\mu\nu}, \phi \right)
f_R\left(R\left({\hat g}_{\mu\nu}, \phi \right)\right) + \kappa^2 T_{\mu\nu}  
\nonumber\\
&&
 + \partial_\mu \phi \partial_\nu \phi
\Big[ 2 f\left(R\left({\hat g}_{\mu\nu}, \phi \right)\right)
 - R\left({\hat g}_{\mu\nu}, \phi \right)
f_R\left(R\left({\hat g}_{\mu\nu}, \phi \right)\right) 
\nonumber\\
&& 
 - 3 \Box\left(g\left({\hat g}_{\mu\nu}, \phi \right)_{\mu\nu}\right)
f_R\left(R\left({\hat g}_{\mu\nu}, \phi \right)\right)+ \kappa^2 T \Big] =0\, ,
\end{eqnarray}
As a convention, here 
  $f_R$ means $\partial f(R)/\partial R$, $\nabla_\mu$ and $\Box$ are  different derivative operators 
  with respect to $g_{\mu\nu}$. Also by using a similar "dictionary"   as we used in $f(R,T)$, we define  
$T_{\mu\nu}$ as the effective matter, energy-momentum tensor for $\mathcal{L}_m$ is given by (\ref{en1}) .
Variation with respect to the scalar field $\phi$ gives us :
\begin{eqnarray}
\label{Eq2}
 &&\!\!\!\!\!\!\!\!\!\!\!\!\!\!
 \nabla\left(g\left({\hat g}_{\mu\nu}, \phi \right)_{\mu\nu}\right)^\mu
\Big\{\partial_\mu \phi \Big[ 2 f\left(R\left({\hat g}_{\mu\nu}, \phi \right)\right)
   - R
\left({\hat g}_{\mu\nu}, 
\phi \right) f_R\left(R\left({\hat g}_{\mu\nu},
\phi \right)\right)
\nonumber\\
&&\ \ \ \ \ \ \ \ \ \ \ \ \ \ \ \ \ \ - 3 \Box\left(g\left({\hat g}_{\mu\nu}, \phi 
\right)_{\mu\nu}\right)
f_R\left(R\left({\hat g}_{\mu\nu}, \phi \right)\right) + \kappa^2 T \Big] \Big\}
=0\, ,
\end{eqnarray}
Here like $f (R, T) $, we define trace of the energy-momentum tensor as $T=g\left({\hat g}_{\mu\nu}, \phi \right)^{\mu\nu}T_{\mu\nu}$ . It has been proven that this new theory 
 is  conformally invariant 
and  ghost free  \cite{Nojiri:2014zqa}. So, it is remarkable to consider it as a valid extension of $f(R)$ gravities.

We consider the same FLRW metric as (\ref{metric}). In this case, the constraint equation leads to
 $\phi=t$ where $t$ is cosmic time. It is not so hard task to write FLRW equations for motion. Actually, because our aim is to investigate Noether symmetry, so what we need is just point like Lagrangian of the  mimetic $f(R)$ scenario. To be more generally speaking, we slightly modify the original mimetic $f(R)$ by including a potential term and by introducing a Lagrange multiplier $\lambda$ as the following:
\begin{equation}
\label{newFRaction}
S=\int d^4 x \sqrt{-g\left(\hat {g}_{\mu\nu}, \phi \right)}
\left[ \frac{f\left(R\left({\hat g}_{\mu\nu}, \phi \right)\right)}{2\kappa^2}
-V(\phi)-
\lambda(\partial_{\mu}\phi\partial^{\mu}\phi+1)+ \mathcal{L}_m\right]\, .
\end{equation}
The point like Lagrangian for a fluid with pressure $p(a,\dot{a},\phi)$ and by taking in to account that $\mathcal{L}_m\equiv p(a,\dot{a},\phi)$, is written in the following form:
\begin{eqnarray}\label{L2}
&&\mathcal{L}(a,R,\phi,\dot{a},\dot{R},\dot{\phi})=a^3 f(R)-a^3R f_{R}(R)-6\Big(a\dot{a}^2f_{R}(R)+a^2\dot{a}\dot{R}f_{RR}(R)\Big)\\&&\nonumber-p(a,\dot{a},\phi)a^3-V(\phi)a^3-\lambda(1-\dot{\phi}^2)a^3
\end{eqnarray}
Note that in general $\lambda=\lambda(t)$. Equations of motion are written in the following forms:
\begin{eqnarray}
&&6f_{RR}\ddot{R}=-6\dot{R}^2f_{RRR}-a\ddot{a}p_{\dot{a}\dot{a}}-\dot{a}
ap_{a\dot{a}}-ap_{\dot{a}t}-3\dot{a}p_{\dot{a}}+ap_a
\\&&\nonumber-f_{R}\Big(-3R+6(\frac{\dot{a}}{a})^2+12\frac{\ddot{a}}{a}-3(f-\lambda+\lambda\dot{\phi}^2-V(\phi)-p)\Big)-12\frac{\dot{a}\dot{R}}{a}
\\&& \frac{\ddot{a}}{a}=\frac{1}{6}(R-6(\frac{\dot{a}}{a})^2)
\\&&\ddot{\phi}+\dot{\phi}(\frac{\dot{\lambda}}{\lambda}+\frac{3\dot{a}}{a})+\frac{1}{2}\frac{V_{\phi}}{\lambda }=0.
\end{eqnarray}
The second equation is just the standard definition of Ricci scalar for FLRW metric. The third one is reduced to the Klein-Gordon equation in the case of $\lambda\equiv \lambda(t)=\texttt{Constant}$. The first equation becomes familiar as the equation of motion in $f(R)$ gravity,  if we set $p=p(a),\lambda=1,\phi(t)=t,V(\phi)=0$. Clearly this equation posses de Sitter solution as $R=R_0$, $a(t)=a_0 e^{H_0 t}$.\par Because of the importance of this model, we will study fixed points of the associated dynamical system, corresponding to this last case , when $p=p(a),\phi=t$. We applied it before in the context of general relativity \cite{Momeni:2009qw}. Here we review the basic concepts of an non-autonomous system.

Consider the following differential equation for an non-autonomous  dynamical state vector $\vec{x}$:
\begin{eqnarray}
&&\dot{\vec{x}}=f(t,\vec{x}) \label{sys1}\\&& f:[0,\infty)\times D\longrightarrow
\mathbb{R}^{n}\\&& D=\{\vec{x} \in\mathbb{R}^{n}   \mid      \parallel
\vec{x}\parallel_{2}<0\}
\end{eqnarray}
The equilibrium point, or fixed point is located at $x=0$ if and only if it solves the following algebraic equation for an instant  of time,namely $t$:
\begin{eqnarray}
&&f(t,0)=0,\forall t\geq 0
\end{eqnarray}
We define the Jacobian matrix ; $J\equiv[\frac{\partial f}{\partial
\vec{x}}]$  must be  bounded  function of  $t$ on a finite domain $\mathcal{D}$ and  furthermore it satisfies smoothly
the Lipschit'z lemma , as the following:
\begin{eqnarray}
&&\parallel f(t,\vec{x})-f(t,\vec{y}) \parallel \leq L \parallel \vec{x}-\vec{y}
\parallel,\parallel \vec{x} \parallel _{p}=(\sum^{p}_{i}\mid x_{i} \mid
^{p})^{\frac{1}{p}},1\leq p <\infty
\end{eqnarray}
There is an important theorem about the asymptotic stability of the system in the vicinity of the equilibrium point:\\
 \texttt{Theorem} I : It is possible to linearize the system of equations in the vicinity of the fixed point in the following form:
\begin{eqnarray}
\dot{\vec{\delta x}}=A(t) \vec{\delta x}\label{sys2}
\end{eqnarray}
  \texttt{Theorem} II: Suppose that $x=0$ be the fixed point of the 
 system $\dot{x}=f(t,x)$  , and it satisfies the following auxiliary conditions:
\begin{eqnarray}
\begin{aligned}
  f:[0,\infty)\times D\longrightarrow
\mathbb{R}^{n}\\ D=\{\vec{x} \in\mathbb{R}^{n} \mid   \parallel
\vec{x}\parallel_{2}<r\},
 \end{aligned}
\end{eqnarray}
  It is adequate to define the time dependent function  $ A(t)=\frac{\partial f(t,x)}{\partial x}|_{x=0}
$ . We say that the system has an  \texttt{exponential stable equilibrium} point of
the linearized  system    (\ref{sys2}) , then this point is the exponential
stable equilibrium point of the  nonlinear system (\ref{sys1})

 In this case the system of equations reduces to the following form:
\begin{eqnarray}
&&6f_{RR}\ddot{R}=-6\dot{R}^2f_{RRR}+ap_a
\\&&\nonumber-f_{R}\Big(-3R+6(\frac{\dot{a}}{a})^2+12\frac{\ddot{a}}{a}-3\Big(f(R)-\lambda+\lambda\dot{\phi}^2-V(\phi)-p\Big)\Big)-12\frac{\dot{a}\dot{R}}{a}
\\&& \frac{\ddot{a}}{a}=\frac{1}{6}(R-6(\frac{\dot{a}}{a})^2)
\\&&\frac{\dot{\lambda}}{\lambda}+\frac{3\dot{a}}{a}+\frac{1}{2}\frac{V_{\phi}}{\lambda }=0.
\end{eqnarray}
The first attempt is done by rewriting the system of equations in terms of a dimensionless "time" coordinate $N=\log a$ and a new set of dimensionless parameters as $x^A=\{h=\frac{H}{H_0},r=\frac{R}{12H_0^2},X=r',\zeta=\log{\lambda}\}$. Equations read as the following non-autonomous system (due to the potential term $V(\phi)=V(N)$:
\begin{eqnarray}
&&h'=2(\frac{r}{h}-h)\\&&
r'=X,\\&&
X'=-\frac{2X}{h}\Big(\frac{r}{h}-h+\frac{6H_0f_{rrr}}{f_{rr}}\Big)\\&&\nonumber-\frac{f_r}{144H_0^4f_{rr}}\Big[-\frac{6r}{h^2}+\frac{4}{h}(\frac{r}{h}-h)-\frac{3}{2H_0^2h^2}(f-V(N)-p(N))\Big]- \frac{X}{72H_0^2 f_{rr}},\\&&
\zeta'=-3-\frac{V(N)}{2H_0}\frac{e^{-\zeta}}{h}.
\end{eqnarray}
Stationary (fixed) points are located at :
\begin{eqnarray}
f(r_c)=V(N_c)+p(N_c)=0,\ \ X_c=0,\ \ r_c=h_c^2,\ \ \frac{V(N_c)}{2H_0}\frac{e^{-\zeta_c}}{h_c}=-3.
\end{eqnarray}
Because of the critical point is function of $N$, it's moving when time is running. The Jacobian of the linearized system is given by the following:
\begin{eqnarray}
J= \left[ \begin {array}{cccc} a_{{1}}&a_{{2}}&0&0\\ \noalign{\medskip}0
&0&1&0\\ \noalign{\medskip}c_{{1}}&c_{{2}}&c_{{3}}&0
\\ \noalign{\medskip}d_{{1}}&0&0&d_{{4}}\end {array} \right]
\end{eqnarray}
Where
\begin{eqnarray}
&&a_i=\frac{\partial f_1}{\partial q^i}, \ \ q^i\equiv\{h,r\},\\&& 
 c_i=\frac{\partial f_3}{\partial q^i},\ \ q^i\equiv\{h,r,X\},\\&& 
d_i=\frac{\partial f_4}{\partial q^i},\ \ q^i=\{h,\zeta\}.
\end{eqnarray}
And:
\begin{eqnarray}
&&f_1=2(\frac{r}{h}-h),\\&&
f_3=-\frac{2X}{h}\Big(\frac{r}{h}-h+\frac{6H_0f_{rrr}}{f_{rr}}\Big)\\&&\nonumber-\frac{f_r}{144H_0^4f_{rr}}\Big[-\frac{6r}{h^2}+\frac{4}{h}(\frac{r}{h}-h)-\frac{3}{2H_0^2h^2}(f-V(N)-p(N))\Big]-\frac{X}{72H_0^2 f_{rr}},\\&&
f_4=-3-\frac{V(N)}{2H_0}\frac{e^{-\zeta}}{h}.
\end{eqnarray}
The characteristic equation is given by:
\begin{eqnarray}
&&{\lambda}^{4}+ \left( -d_{{4}}-c_{{3}}-a_{{1}} \right) {\lambda}^{3}+
 \left( d_{{4}}c_{{3}}+d_{{4}}a_{{1}}-c_{{2}}+c_{{3}}a_{{1}} \right) {
\lambda}^{2}\\&&\nonumber+ \left( d_{{4}}c_{{2}}-d_{{4}}c_{{3}}a_{{1}}-c_{{1}}a_{{2
}}+c_{{2}}a_{{1}} \right) \lambda-d_{{4}} \left( -c_{{1}}a_{{2}}+c_{{2
}}a_{{1}} \right)=0.
\end{eqnarray}
one eigenvalue is $\lambda_1=d_4$. Another eigenvalues read as the following:
\begin{eqnarray}
&&\lambda_2=\,{\frac {{\Delta}^{2}+12\,c_{{2}}-4\,c_{{3}}a_{{1}}+4\,{c_{{3}}}^{
2}+4\,{a_{{1}}}^{2}+2\,c_{{3}}\Delta+2\,a_{{1}}\Delta}{6\Delta}}
\end{eqnarray}
\begin{eqnarray}
&&\lambda_3=-\frac{1}{12}\,\chi-{\frac {c_{{2}}}{\chi}}+\frac{1}{3}\,{\frac {c_{{3}}a_{{1}}}{\chi}
}-\frac{1}{3}\,{\frac {{c_{{3}}}^{2}}{\chi}}-\frac{1}{3}\,{\frac {{a_{{1}}}^{2}}{\chi}
}+\frac{1}{3}\,c_{{3}}+\frac{1}{3}\,a_{{1}}\\&&\nonumber+\frac{1}{12}\,i\sqrt {3}\chi-{\frac {i\sqrt {3}c_{
{2}}}{\chi}}+{\frac {\frac{1}{3}\,i\sqrt {3}c_{{3}}a_{{1}}}{\chi}}-{\frac {\frac{1}{3}
\,i\sqrt {3}{c_{{3}}}^{2}}{\chi}}-{\frac {\frac{1}{3}\,i\sqrt {3}{a_{{1}}}^{2}
}{\chi}}
\end{eqnarray}
\begin{eqnarray}
&&\lambda_4=-\lambda_3^{*}.
\end{eqnarray}
In the above expressions :
\begin{eqnarray}
&&\Delta\equiv 12 \it E+8\,{a_{{1}}}^{3}-12\,c_{{3}}{a_{{1}}}^
{2}+ \left( -12\,{c_{{3}}}^{2}-72\,c_{{2}} \right) a_{{1}}+36\,c_{{2}}
c_{{3}}+8\,{c_{{3}}}^{3}+108\,c_{{1}}a_{{2}},\\&&
E^2\equiv54\,c_{{2}}c_{{3}}c_{{1}}a_{{2}}-108\,c_{{2}}a_{{1}}c_{{1}}a_{{2}}-18
\,{c_{{3}}}^{2}a_{{1}}c_{{1}}a_{{2}}-18\,c_{{3}}{a_{{1}}}^{2}c_{{1}}a_
{{2}}-12\,{c_{{2}}}^{3}+81\,{c_{{1}}}^{2}{a_{{2}}}^{2}-3\,{c_{{2}}}^{2
}{c_{{3}}}^{2}\\&&\nonumber+24\,{c_{{2}}}^{2}{a_{{1}}}^{2}-12\,c_{{2}}{a_{{1}}}^{4}
+6\,{c_{{3}}}^{3}{a_{{1}}}^{3}-3\,{c_{{3}}}^{4}{a_{{1}}}^{2}-3\,{c_{{3
}}}^{2}{a_{{1}}}^{4}+12\,c_{{1}}a_{{2}}{c_{{3}}}^{3}+12\,c_{{1}}a_{{2}
}{a_{{1}}}^{3}-24\,{c_{{2}}}^{2}c_{{3}}a_{{1}}\\&&\nonumber-6\,c_{{2}}{c_{{3}}}^{2}
{a_{{1}}}^{2}-6\,c_{{2}}{c_{{3}}}^{3}a_{{1}}+24\,c_{{2}}c_{{3}}{a_{{1}
}}^{3}\\&&
{\chi}^{3}=36\,c_{{2}}c_{{3}}-72\,c_{{2}}a_{{1}}-12\,{c_{{3}}}^{2}a_{{1}}-12\,c_{
{3}}{a_{{1}}}^{2}+108\,c_{{1}}a_{{2}}+8\,{c_{{3}}}^{3}+8\,{a_{{1}}}^{3
}+12\,\rho,\\&&
{\rho}^{2}=
54\,c_{{2}}c_{{3}}c_{{1}}a_{{2}}-108\,c_{{2}}a_{{1}}c_{{1}}a_{{2}}-18
\,{c_{{3}}}^{2}a_{{1}}c_{{1}}a_{{2}}-18\,c_{{3}}{a_{{1}}}^{2}c_{{1}}a_
{{2}}+81\,{c_{{1}}}^{2}{a_{{2}}}^{2}-3\,{c_{{2}}}^{2}{c_{{3}}}^{2}+24
\,{c_{{2}}}^{2}{a_{{1}}}^{2}\\&&\nonumber-12\,c_{{2}}{a_{{1}}}^{4}+6\,{c_{{3}}}^{3}
{a_{{1}}}^{3}-3\,{c_{{3}}}^{4}{a_{{1}}}^{2}-3\,{c_{{3}}}^{2}{a_{{1}}}^
{4}-12\,{c_{{2}}}^{3}+12\,c_{{1}}a_{{2}}{c_{{3}}}^{3}+12\,c_{{1}}a_{{2
}}{a_{{1}}}^{3}-24\,{c_{{2}}}^{2}c_{{3}}a_{{1}}-6\,c_{{2}}{c_{{3}}}^{2
}{a_{{1}}}^{2}\\&&\nonumber-6\,c_{{2}}{c_{{3}}}^{3}a_{{1}}+24\,c_{{2}}c_{{3}}{a_{{1
}}}^{3}.
\end{eqnarray}
To have stable solution we must have $d_4=-3+\frac{V(N_c)}{2H_0}\frac{e^{-\zeta_c}}{h_c}<0$. Another eigenvalues $\lambda_i,\ \ 2\leq i\leq 4$ must satisfy $\Re{\lambda_i}<0$. It is remarkable to mention here that the stable manifold is defineb by the de Sitter space-time. So, the model for time intervals $N>N_c$ has stable de Sitter solution.

\section{\label{NS} Noether symmetry approach to dynamical systems}
Symmetry is an important issue to be addressed in any physical theory. The basic property of a system with a defenite type of symmetry is the existence of an associated conserved quantity under this kind of symmetry. Let us consider a typical dynamical system, is defined by a set of configurations coordinates set  $q_i$. Generally speaking, the dimension of the system is defined as the number of independent coordinates of the system. We assume that the dynamics of the system are defined by the point like  Lagrangian  is given by:$L\equiv L(q_i, \dot{q}_i;t),\ \ 1\leq  i\leq N$.
For each coordinate, it is possible to define a "unique" first order conjugate momentum:
\begin{eqnarray}
p_{i}\equiv\frac{\partial L}{\partial \dot{q}^i}.
\end{eqnarray}
Euler-Lagrange equation of motion is given by the following set of N-ordinary second order diffrential equations:
\begin{eqnarray}
\dot{p}_i-\frac{\partial L}{\partial q^{i}}=0,
\end{eqnarray} 
What we call it as {\it Noether Symmetry Approach}\cite{noether1}-\cite{noether4} is 
the existence of a "unique" vector field  $\vec{X}$ on tangent space  $T{\cal Q}\equiv\{q_i, \dot{q}_i\}$): :

\begin{equation}\label{17}
\vec{X}=\Sigma_{i=1}^{N}\Big[\alpha^i(q)\frac{\partial}{\partial q^i}+
 \dot{\alpha}^i(q)\frac{\partial}{\partial\dot{q}^i}\Big]\,{,}
 \end{equation}
If we can find "generators" coefficients 
 $\alpha_{i}(q_j)$, then we can strictly say that our dynamical system must satisfy the following geometrical constraint (is called as {\it Lie derivative of the Lagrangian} :
 \begin{equation}\label{19}
 L_X{\cal L}=0\,
 \end{equation}
where explicitly we have:
 \begin{equation}\label{ns}
 L_X{\cal L}=\vec{X}{\cal L}=\Sigma_{i=1}^{N}\Big[\alpha^i(q)\frac{\partial {\cal L}}{\partial q^i}+
 \dot{\alpha}^i(q)\frac{\partial {\cal L}}{\partial\dot{q}^i}\Big]\,{.}
 \end{equation}
or equivalently we can write it as the following:
 \begin{equation}\label{21}
 \Sigma_{i=1}^{N}\frac{d}{dt}\left(\alpha^i\frac{\partial {\cal
 L}}{\partial\dot{q}^i}\right)=L_X{\cal L}\,{.}
 \end{equation}
It is an easy task to show that, existence of Noether symmetry implies that the system has the following conserved (local) quantity:
 \begin{equation}\label{22}
 \Sigma_0=\Sigma_{i=1}^{N}\alpha^ip_i
 \end{equation}
The equations are obtained setting to zero the coefficients of the terms $\dot{q_i}^m\dot{q}_j^n, 0\leq m+n\leq N$ in (\ref{ns}).
They are several interesting applications of this symmetry approach in different 
cosmological models in different models \cite{cap3}-\cite{Jamil:2011pv}.  In our paper we'll apply this approach to $f(R,T)$ and  mimetic $f(R)$ theories of gravity.

\section{Noether symmetry  in $f(R,T)$ gravity}\label{NS1}
The plan in this section is to study the system of Noether equations for Lagrangian given by (\ref{L1}). If we write down (\ref{ns})  equation for (\ref{L1}) ,
\begin{eqnarray}
&&\alpha\frac{\partial \mathcal{L}}{\partial a}+\beta\frac{\partial \mathcal{L}}{\partial R}+\gamma\frac{\partial \mathcal{L}}{\partial T}+\Big(\dot{a}\frac{\partial\alpha}{\partial a}+\dot{R}\frac{\partial \alpha}{\partial R}+\dot{T}\frac{\partial \alpha}{\partial T}\Big)\frac{\partial \mathcal{L}}{\partial \dot{a}}\\&&\nonumber+\Big(\dot{a}\frac{\partial \beta}{\partial a}+\dot{R}\frac{\partial \mathcal{\beta}}{\partial R}+\dot{T}\frac{\partial \mathcal{\beta}}{\partial T}\Big)\frac{\partial \mathcal{L}}{\partial \dot{R}}+\Big(\dot{a}\frac{\partial \gamma}{\partial a}+\dot{R}\frac{\partial \gamma}{\partial R}+\dot{T}\frac{\partial \gamma}{\partial T}\Big)\frac{\partial \mathcal{L}}{\partial \dot{T}}=0.
\end{eqnarray}
and by putting the coefficients of $\{\dot{a}^2,\dot{R}^2,\dot{T}^2,\dot{a}\dot{R},\dot{a}\dot{T},\dot{T}\dot{R}\}$ and constant terms, we obtain the following system of partial differential equations for $\vec{X}$ components $\{\alpha,\beta,\gamma\}$:

\begin{eqnarray}
\label{eq1}&&\dot{a}^2:\ \ \alpha f_{R}+\beta a f_{RR}+\gamma a f_{RT}+2a\alpha_{a}f_{R}+a^2f_{RR}\beta_{a}+\gamma_a a^2 f_{RT}=0,\\
\label{eq2}&& f_{RR}\alpha_{R}=0,\ \ f_{RT}\alpha_{T}=0,\ \ \alpha_{a} f_{RR}=0\\ &&
\label{eq3}\dot{a}\dot{R}:\ \ 
a_{{}} f_{RR}  \beta+a_{{}} \gamma f_{TTT}
+a_{{}}f _{RR}
\alpha _a +a_{{}} \beta_R
  f _{RR}\\&&\nonumber +a_{{}} \gamma{R}f_{RT} +2\alpha _R  f _R +2\,
 f _{RR}  \alpha =0
,\\&&
\label{eq4}\dot{a}\dot{T}: \ \ a_{{}} f_{RRR} \beta  +a_{{}} f_{RRT} \gamma+ \left( 2\,\alpha +a_{{}}
\alpha _a +a_{{}}\beta _R \right) f_{RR}+a_{{}}\gamma_R f_{RT} +2\alpha_Rf _R =0
\label{eq5}\\&&\dot{R}\dot{T}: \alpha_R f_{RT}+\alpha_Tf_{RR}=0\ \ 
\\&&
\label{eq6} \Big( \gamma R_{{}}+\beta
   \left( T_{{}}-3\,p \left( a_{{}}
 \right) +\rho \left( a_{{}} \right)  \right)  \Big) a_{{}}f_{RT}\\&&\nonumber +\gamma  a_{{}} \Big( T_{
{}}-3\,p \left( a_{{}} \right) +\rho \left( a_{{}} \right)  \Big)f_{TT} +\beta a_{{}}R_{{}}
f_{RR}\\&&\nonumber +3\Big( ( T_{{}}-3p (a) +\rho
 (a) +\frac{1}{3}a\rho'(a) - ap'(a)\Big)f_{T} -f  +\frac{1}{3}
 ap'(a)+p (a) +Rf_{R} \alpha  =0.
\end{eqnarray}
\\
It is a hard job to find all possible solutions of this nonlinear system of first order coupled partial differential equations (PDEs). Inspired directly from the case of general relativity we'll limit ourselves to the following simple cases:
\begin{itemize}
  \item The case of Einstein gravity with matter components $f(R,T)=R+2\Lambda+g(T)$: because any theory of  modified gravity must be reduced to Einstein-Hilbert action  at low curvature regime, we are interested to study a solution of $f(R,T)$ in which the Einstein-Hilbert term is dominated as the leading term of theory. If we substitute $f(R,T)=R+2\Lambda+g(T)$, we obtain the following
exact solution:
\begin{eqnarray}
&&\alpha=\frac{C_1}{\sqrt{a}},\ \ \beta=\gamma=0\\&&
g(T)=C_2e^{-3C_1\int{\frac{dT}{\gamma(T)}}}\\&&
p(a)=2\Lambda+\frac{C_1}{a^3}.
\end{eqnarray}
We mention here that this is an special class of solutions founded before in literature \cite{Capozziello:2008ch} (for the case of purely $f(R,T)=F(R)$ see \cite{Capozziello:2008ch}). The associated Noether charge reads:
\begin{eqnarray}
-12C_1\dot{a}\sqrt{a}=\Sigma_0.
\end{eqnarray}
The corresponding scale factor is obtained as follows:
\begin{eqnarray}
a= (\frac{3A}{2})^{2/3}\Big[1-\frac{1}{3}H_0 t\Big]^{2/3},\ \ H_0=\frac{3\Sigma_0}{8C_1}.
\end{eqnarray}
This solution can be written in terms of q-exponential family \cite{Setare:2014vna}
:
\begin{eqnarray}
a(t)=a_0 e_{2/3}(H_0 t),\ \ a_0=(\frac{3A}{2})^{2/3}.
\end{eqnarray}
The pressure term deserves more investigations. The first term implies on the existence of the "background pressure" $p_0=2\Lambda$, which it can be realized by expectation vacuum energy of some quantum fields. The second term is a dark matter term, if we identify $c_1=\rho_{m0}$, as the dark matter density at present era, $t=0,a_0=1$.
In the case of perfect fluid with equation of state $w=\frac{p}{\rho}$, we have $\rho\sim a^{-3(1+w)}$. By a power-law expansion, $a(t)\sim t^p$, to be accelerated Universe we must have $p>1$, in our case the solution is not accelerating solution.

\item Case with $f(R,T)=h(R)+k(T)$: if we would like to pass to modified gravity, we should try to solve systems of equations by assuming that $R\to h(R)$, and by taking into the account the matter sector $k(T)$. This simple assumption gives us the following exact solutions for the model (\ref{L1}) under Noether symmetry approach:
\begin{eqnarray}
&&\alpha=0,\ \ \beta=\frac{F_{1}(R)}{a},\ \ \gamma=\frac{R}{a}\Delta(T),\\&&
h(R)=C_1\int_{0}^{R} (R-\tau)F_{1}(\tau)d\tau,\ \ k(T)=C_1\int{\frac{dT}{\Delta(T)}}+C_2.
\end{eqnarray}
here $\{\Delta(T),F_{1}(R)\}$ stands for a pair of arbitrary functions.
The conserved Noether charge associated with this model reads as follows:
\begin{eqnarray}
\sigma_0=-6C_1 a\dot{a}F_{1}(R)^2.
\end{eqnarray}
 Due to the leakage of more information of the form of $F_1(x)$, we cannot integrate it. But it is possible to solve it for Starobinsky inflationary model $F(R)=R+\frac{R^2}{6M^2}$ solution as the following:
\begin{eqnarray}
&&t+C_2=\int_{0}^{a(t)}{\frac{xdx}{R(-x^4-24 j(y))+4C_1}}
\end{eqnarray}
here
\begin{eqnarray}
 R(-x^4-24j(y))=Root\Big(-x^4-24j(y)=0|_{y=x}\Big),\end{eqnarray}
and we define 
\begin{eqnarray}
&&j(y)\equiv\frac{1}{27}\,{\frac { \left( 9\,{C_{{1}}}^{2}{y}^{2}{M}^{2}-C_{{1}}y\sigma_{{0
}} \right) ^{3/2}}{\sigma_{{0}}{M}^{3}{C_{{1}}}^{2}}}+\frac{1}{36}\,{\frac {y
\sqrt {9\,{C_{{1}}}^{2}{y}^{2}{M}^{2}-C_{{1}}y\sigma_{{0}}}}{{M}^{3}C_
{{1}}}}\\&&\nonumber-{\frac {1}{648}}\,{\frac {\sigma_{{0}}\sqrt {9\,{C_{{1}}}^{2}{
y}^{2}{M}^{2}-C_{{1}}y\sigma_{{0}}}}{{M}^{5}{C_{{1}}}^{2}}}
-\frac {\sqrt {9}\sigma_{0}^{2}}
{11664C_1^2M^6}\ln \Big ( \frac { ( -\frac{1}{2}C_{
1}\sigma_{0}+9y^2C_{1}^{2}M^{2} )}{9C_1M}\Big) \\&&\nonumber+\sqrt {9}{M}^{-6}C_{1}^{-2}\sqrt {9C_{1}^{2}y^{2}M^{2}-C_1y
\sigma_0} ) +\frac {y^{3}C_{1}}{\sigma_{0}}.
\end{eqnarray}

\end{itemize}

\section{Noether symmetry for  mimetic $f(R)$ gravity}\label{NS2}
In this section we'll see how Noether symmetry gives us useful information about  mimetic $f (R) $ theory. Especially we would like to search for the possible forms of potential function $V(\phi)$ in this model. Thanks to the normalization condition $\partial_{\mu}\phi\partial^{\mu}\phi=1$, the potential is an implicit function of $t$. So, generally speak in the dynamical system Lagrangian is time dependent. Let us to start by the same method as we used in the previous section. 
We write the following condition of (\ref{ns}) for $f(R)$ mimetic model:
\begin{eqnarray}
&&\alpha\frac{\partial \mathcal{L}}{\partial a}+\beta\frac{\partial \mathcal{L}}{\partial R}+\gamma\frac{\partial \mathcal{L}}{\partial \phi}+\Big(\dot{a}\frac{\partial\alpha}{\partial a}+\dot{R}\frac{\partial \alpha}{\partial R}+\dot{\phi}\frac{\partial \alpha}{\partial T}\Big)\frac{\partial \mathcal{L}}{\partial \dot{a}}\\&&\nonumber+\Big(\dot{a}\frac{\partial \beta}{\partial a}+\dot{R}\frac{\partial \mathcal{\beta}}{\partial R}+\dot{\phi}\frac{\partial \mathcal{\beta}}{\partial \phi}\Big)\frac{\partial \mathcal{L}}{\partial \dot{R}}+\Big(\dot{a}\frac{\partial \gamma}{\partial a}+\dot{R}\frac{\partial \gamma}{\partial R}+\dot{\phi}\frac{\partial \gamma}{\partial \phi}\Big)\frac{\partial \mathcal{L}}{\partial \dot{\phi}}=0.
\end{eqnarray}

where for simplicity we assume that $\lambda=\lambda(t),p=p(a)$. Using (\ref{L2}), we can write the following system of differential equations , linear in $\{\alpha,\beta,\gamma\}$:
\begin{eqnarray}
&&-3\alpha Rf_R-\beta aRf_{RR}+3\alpha  f-3\alpha\lambda -3\alpha V(\phi)-\gamma a V_{\phi}-\alpha a p'(a)-3\alpha p =0\\&&
3\alpha  +2 a\gamma_{\phi}=0\\&&
2\alpha_{a}af_{R}+\beta a f_{RR}+a^2\beta_a f_{RR}+\alpha f_{R}=0\\&&
\alpha_R  f_{RR}=0\\&&
\lambda a^2 \gamma_a-6\alpha_{\phi}f_R-3a\beta_{\phi}
f_{RR}=0\\&&
2a\alpha_Rf_{R}+2a\alpha f_{RR}+\beta a^2f_{RRR}+a^2\alpha_a f_{RR}+a^2\beta_{R}f_{RR}=0\\&&
\lambda a \gamma_R-3\alpha_{\phi}f_{RR}=0.
\end{eqnarray}
which are obtained setting to zero the coefficients of the different terms $\{\dot{a}^2,\dot{R}^2,\dot{\phi}^2,\dot{a}\dot{R},\dot{a}\dot{\phi},\dot{\phi}\dot{R}\}$.
We can distinguish some possible cases:

 \begin{itemize}
\item  Case of GR: the possible non trivial solution for $f(R)=R+2\Lambda$ read as the following:

\begin{eqnarray}
&& V \left( \phi \right) ={\it C_4}\\&&
\alpha \left( a,R,\phi \right) =\frac{1}{3}\,
{a}^{1+{\it 
c}_{{1}}}{\it C_1}\sqrt { c_1\lambda}\left({\it C_3}\,\sin \left( \frac{\sqrt { c_1\lambda}}{2}
\phi \right) -
{\it C_2}\,\cos \left( \frac{\sqrt { c_1\lambda}}{2}
\phi \right)   \right) ,\\&&
\beta \left( a,R,\phi \right) =-
\frac{\left( 2\,{\it 
c}_{{1}}+3 \right){\it C_1}\,{a}^{{\it c}_{{1}}}\sqrt { c_1\lambda}}
{3R}\left( {\it C_3}\,\sin \left( \frac{\sqrt { c_1\lambda}}{2}\phi \right) -{\it C_2}\,\cos \left( \frac{\sqrt { c_1\lambda}}{2}\phi \right)  \right)  ,\\&&
\gamma  \left( a,R,\phi
 \right) ={\it C_1}\,{a}^{{\it c}_{{1}}} \left( {\it C_2}\,\sin
 \left( \frac{\sqrt { c_1\lambda}}{2}\phi \right) +{\it 
C_3}\,\cos \left( \frac{\sqrt { c_1\lambda}}{2}\phi
 \right)  \right) ,\\&&
p \left( a \right) =2\,\Lambda-
\lambda-{\it C_4}+{a}^{3}{\it C_5} . 
\end{eqnarray}
This solution corresponds to a constant potential form. The Noether conserved charge reads as the following:
\begin{eqnarray}
&&Q=-4\dot{a}\,{a}^{2+{\it 
c}_{{1}}}{\it C_1}\sqrt { c_1\lambda}\left({\it C_3}\,\sin \left( \frac{\sqrt { c_1\lambda}}{2}
\phi \right) -
{\it C_2}\,\cos \left( \frac{\sqrt { c_1\lambda}}{2}
\phi \right)   \right)\\&&\nonumber+2\lambda{\it C_1}\,{a}^{{\it c}_{{1}}+3} \left( {\it C_2}\,\sin
 \left( \frac{\sqrt { c_1\lambda}}{2}\phi \right) +{\it 
C_3}\,\cos \left( \frac{\sqrt { c_1\lambda}}{2}\phi
 \right)  \right) 
\end{eqnarray}
Because $\phi=t$ the equation is integrable to give us $a(t)$. For $Q=0$, the exact solution existed for the scale factor:
\begin{eqnarray}
&&a(t)=a_0\left({\it C_3}\,\sin \left( \omega
t \right) -
{\it C_2}\,\cos \left(\omega
t\right)   \right) ^{n},\ \ n=\frac{1}{c_1} ,\ \  \omega=\frac{\sqrt { c_1\lambda}}{2}\label{a}.
\end{eqnarray}
It defines an oscillatory solution  with Type IV future singularity \cite{Nojiri:2005sr}. If $C_2=0,\omega\to i \omega,n=\frac{3}{2}$,
It is identified in the late-time $\Lambda$CDM era.

 Following  \cite{singularities}, we can classify the future singularities as follow:
\begin{itemize}
\item TypeI: ("BigRip"):   $t\to t_s$,$a\to \infty,\rho\to \infty$ and $|p| \to \infty$. 

\item  TypeII: ("sudden"):  
$t\to t_s$,$a\to a_s,\rho\to \rho_s$ and $|p| \to \infty$. 

\item  Type III : 
$t\to t_s$,$a\to a_s,\rho\to \infty$ and $|p| \to \infty$

\item  Type IV : 
$t\to t_s$,$a\to a_s,\rho\to 0$ and $|p| \to 0$
 and
higher derivatives of H diverge.
 \par
Here $t_s, a_s$ and$\rho_s$ are constants with $a_s\neq0$.

\end{itemize}
For our scale factor (\ref{a}) the Hubble parameter and first and second derivatives of $H$ read as the following:
\begin{eqnarray}
&&H(t)=\frac{\dot{a}}{a}= n\omega\Big[\frac{C_3+C_2\tan(\omega t)}{C_3\tan(\omega t)-C_2} \Big] \\&&
\dot{H}= -n\omega^2(C_2^2+C_3^2)\sec^2(\omega t)(C_3\tan(\omega t)-C_2) ^{-2}\\&&
\ddot{H}= \frac {2n{\omega}^{3} ( {C_{{2}}}^{2}+{C_{{3}}}^{2}) 
 ( \sin( \omega\,t ) C_{{2}}+C_{{3}}\cos( 
\omega\,t )) }{C_3^3\sin^3(\omega t)-C_2^3\cos^3(\omega t)-3C_3C_2\Big(C_3\sin(\omega t)-C_2\cos(\omega t)\Big)}
\end{eqnarray}

There is solution(s) for $H(t_s)=\infty$ where
\begin{eqnarray}
t_s^{n}=n\pi\pm\arctan\Big[\frac{C_{2,3}}{C_2^2+C_3^2}\Big],\ \ n\in \mathcal{Z}.
\end{eqnarray}
Where
\begin{eqnarray}
\lim_{t\to t_s} H^{(n\geq 2)}\to \infty.
\end{eqnarray}
So, our solution represents Type IV future singularities. It is remarkable that scale factor remains finite at $t\to t_s$.  Also, because we must have
\begin{eqnarray}
\lim_{t\to t_s} p \left( a \right) =\lim _{a\to a_s}\Big(2\,\Lambda-
\lambda-{\it C_4}+{a}^{3}{\it C_5}\Big)=0
\end{eqnarray}
we obtain:
\begin{eqnarray}
2\,\Lambda-
\lambda-{\it C_4}+{a_s}^{3}{\it C_5}=0
\end{eqnarray}
the cosmological constant term is "calibrated" as the following:
\begin{eqnarray}
\Lambda=\frac{1}{2}(
\lambda+{\it C_4}-{a_s}^{3}{\it C_5}).
\end{eqnarray}

\item  Modified gravity : if we put $f(R)=R+2\Lambda+h(R)$ a very careful analysis of the system of equations gives us the following solutions:
\subsection{Solution with quadratic potential form: Hybrid inflation model}
The system of equtions has the following exact solutions for a set of  functions:
\begin{eqnarray}
&& V \left( \phi \right) ={\it C_5}+{\it C_6}\, \left( \phi+{
\frac {{\it C_2}}{{\it C_1}}} \right) ^{2},\\&&
\alpha \left( a,\phi
 \right) =-\frac{2}{3}\,a{\it C_1},\\&&
\beta \left( a,R,\phi \right) =2\,{\frac {
{\it C_1}}{R}},\\&&
\gamma \left( a,R,\phi \right) ={\it C_1}\,\phi+{\it 
C_2},\\&&
h \left( R \right) ={\it C_3}\,R+{\it C_4},\\&&
p \left( a \right) 
=-{\it C_5}+2\,\Lambda+{\it C_4}-\lambda+{a}^{3}{\it C_7} .
\end{eqnarray}
The quadratic potential is for a massive scalar. This type of potential used as an inflationary model, so called as a Hybrid inflation model. More precisely, if we set $C_2=0$, $C_5=V_0,C_6=\frac{m^2}{2}$, the potential in this regime is written as:
\begin{eqnarray}
V(\phi)=V_0+\frac{m^2}{2}\phi^2.
\end{eqnarray}
This case is a vacuum energy, drive inflation, and is  distinct from the  other classes, like large-field  or small-field models \cite{J.Yokoyama}. Conserved charge is written as the following:
\begin{eqnarray}
\Sigma^{*}=8C_1C_3a^2\dot{a}+2\lambda a^3(C_1 t+C_2).
\end{eqnarray}

Specially by  integrating this equation for $\Sigma^{*}=0$ we obtain:
\begin{eqnarray}
&&a(t)=a_0{{\rm e}^{-\,{\frac {\lambda\,t \left( C_{{1}}t+2\,C_{{2}} \right) 
}{8C_{{1}}C_{{3}}}}}}.
\end{eqnarray}
This model  is described as  the bouncing universe \cite{bounce1,bounce2} $a(t)\sim e^{\alpha t^2}$. To show this equivalence, we rewrite it in the following form:
\begin{eqnarray}
&&a(t)=\tilde{a}_0 e^{-\frac{\lambda C_2^2}{8C_3C_1^2}t^2},\ \ \tilde{a}_0\equiv a_0e^{\frac{\lambda C_2^2}{C_3C_1^2}}.
\end{eqnarray}
Hubble parameter and deceleration parameter read as the following:
\begin{eqnarray}
H=-\frac{\lambda C_2^2}{4C_3C_1^2}t,\ \ q=-(1+\frac{\dot{H}}{H^2})=-(1-\frac{4C_3C_1^2}{\lambda C_2^2t^2}).
\end{eqnarray}
To have acceleration expansion we should have:
\begin{eqnarray}
-\frac{\lambda C_2^2}{4C_3C_1^2}(1-\frac{\lambda C_2^2t^2}{4C_3C_1^2})>0.
\end{eqnarray}
We obtain the following cases:
\begin{itemize}
\item  If $\frac{\lambda}{C_3}>0$ then we have acceleration for time intervals $t\in(-\infty,|\frac{4C_3}{\lambda}|^{1/2}\frac{C_1}{C_2})$.
\item  If $\frac{\lambda}{C_3}<0$ then we have acceleration for time intervals $t\in(|\frac{4C_3}{\lambda}|^{1/2}\frac{C_1}{C_2},\infty)$.
\end{itemize}
In the above discussions, we only consider the case that the universe is expanding.
\subsection{Solution with exponential inflationary models}
A possible solution is the following hyperbolic model:
\begin{eqnarray}
&&V \left( \phi \right) = {\it C_6}+2{\it C_7} \cosh{\frac {2(
\phi+ C_2)}{ C_1}} \\&&
\alpha \left( a,\phi \right) =\frac{2}{3C_1}\,i{a}^{1+{\it C_4}}\sqrt {{\it 
C_3}} \sinh{\frac {2(
\phi+ C_2)}{ C_1}}
,\\&&
\gamma  \left( a,R,\phi \right) =-\,
i{a}^{{\it C_4}}
\sqrt {{\it C_3}} \cosh{\frac {2(
\phi+ C_2)}{ C_1}} ,\\&&
\beta \left( a,R,\phi \right) =-2i\frac{\sqrt {{
\it C_3}} \left( 2\,{\it C_4}+3 \right) \,{a}^{{\it 
C_4}}}{3{{\it C_1}}{R}} \sinh{\frac {2(
\phi+ C_2)}{ C_1}}
,\\&&
h
 \left( R \right) =-(\frac{1}{4}\,\lambda\,{\it C_4}\,{{\it C_1}}^{2}+1)R+{
\it C_5},\\&&
p \left( a \right) = - 
(\lambda +{\it C_6}+6\Lambda
+6{\it C_7}+3{\it C_5})+
 C_8 a^3
\end{eqnarray}
Because all functions must be real valued, so it is adequate to write $C_3=-M^2$, where $M$ is a free parameter. The associated conserved charge reads:
\begin{eqnarray}
Q_0=a^{2+C_4}\Big[\dot{a}\frac{2(4-C_4 C_1^2)M}{C_1}\sinh{\frac {2(
t+ C_2)}{ C_1}}+2M\lambda a\cosh{\frac {2(
t+ C_2)}{ C_1}}\Big]
\end{eqnarray}
It is completely integrable,specially when $Q_0=0$, we obtain:
\begin{eqnarray}
&&a(t)={\it C_1}\, \left( \sinh \left( 2\,{\frac {t+C_{{
2}}}{C_{{1}}}} \right)  \right) ^{n},\ \ n=\,{\frac {\lambda\,{C_{{1}}}^{2}}
{2(-4+C_{{4}}{C_{{1}}}^{2})}}.
\end{eqnarray}
In case of $n=\frac{3}{2}$ it coincides with the late-time $\Lambda$CDM era.
\end{itemize}


\section{Conclusions}
Motivated by recent observational data, indicates that we live in an accelerating universe, several forms of modified gravities have been proposed to resolve and explain this physical phenomena. One of the most popular and physically acceptable candidates is $f(R)$ gravity and its extensions. In our work we established Noether symmetry issue for two types of $f(R)$ theories: a type of non-minimally coupled model is called as $f(R,T)$ and mimetic $f(R)$. We started by reviewing the basic physical foundations of these theories. In $f(R,T)$ model we have been written point-like Lagrangian for flat FLRW metric. We studied equations of motion and Noether symmetry form for it. Two important classes of solution for $f(R,T)$ were found. In the first class, we show that the generalized q-exponential scale factor is an exact solution which it mimics the background with the background pressure. Other solutions were found as general family of additive models, with an exact solution for scale factor in terms of elementary functions. This cosmological solution was obtained by considering the  Starobinsky model $f(R)=R+\alpha R^2$. In mimetic $f(R)$ theory, we have been considered Noether symmetries. We observed that there are two classes of solutions: the first is equivalent to the GR with dark matter and the calibrated cosmological constant. This case mimics a type of cosmological solutions with type IV future singularities, where higher derivatives of $H$ diverge. Another case is modified gravity with two specified forms of potential functions: hybrid inflationary model,in which the scale factor evolves in the bouncing scenario. The second family is exponential form, in this case scale factor mimics the form of $\Lambda CDM$ model perfectly. So, all cosmological models including  $\Lambda CDM$,bouncing and oscillatory solutions with future singularities are described perfectly by Noether symmetrized $f(R,T)$ and mimetic $f(R)$ theories. We conclude that Noether symmetry is able to provide a very excellent way to study cosmological implications of extended $f(R)$ theories.

\section*{Acknowledgments}
We would like to thank the anonymous reviewer for enlightening comments related
 to this work.


\end{document}